# Gas adsorption on MoS$_2$ monolayer from first-principles calculations

Shijun Zhao,[1,2] Jianming Xue[1,2,*] and Wei Kang[2]

[1]State Key Laboratory of Nuclear Physics and Technology, School of Physics, Peking University, Beijing 100871, P. R. China

[2]Center for Applied Physics and Technology, Peking University, Beijing 100871, P. R. China

**Abstract:**

First-principles calculations within density functional theory (DFT) have been carried out to investigate the adsorption of various gas molecules including CO, CO$_2$, NH$_3$, NO and NO$_2$ on MoS$_2$ monolayer in order to fully exploit the gas sensing capabilities of MoS$_2$. By including van der Waals (vdW) interactions between gas molecules and MoS$_2$, we find that only NO and NO$_2$ can bind strongly to MoS$_2$ sheet with large adsorption energies, which is in line with experimental observations. The charge transfer and the variation of electronic structures are discussed in view of the density of states and molecular orbitals of the gas molecules. Our results thus provide a theoretical basis for the potential applications of MoS$_2$ monolayer in gas sensing and give an explanation for recent experimental findings.

Corresponding author: jmxue@pku.edu.cn

## I. Introduction

Two-dimensional (2D) nanomaterials such as graphene and hexagonal BN sheet have attracted great interest because of their extraordinary properties attributed to their ultrathin thickness and related quantum effects.[1-3] Recently, several new classes of 2D nanostructures like transition metal dichalcogenides (TMDs) have been realized experimentally.[4-7] Specifically, suspended single-layer molybdenum disulfide ($MoS_2$) sheet (1H-$MoS_2$) has been successfully synthesized and its unique characteristics distinguished from corresponding bulk counterpart have been demonstrated.[2, 8-10] With one Mo and two S atoms positioned at alternating corners in hexagons, 1H-$MoS_2$ hold tremendous promise for many applications as a result of its exceptional catalytic[11], photovoltaic[9] and lubricant[12] characteristic. In particular, $MoS_2$ based field-effect transistors and sensing films had experimentally shown gas sensing capabilities for NO and $NH_3$ with a ultrahigh sensitivity.[13, 14]

Detection of gas molecules especially toxic gas is extremely important and critical to industry and public health. When monolayer $MoS_2$ is used as gas sensors, its sensor properties rely on the changes in the resistivity due to molecules adsorbed on the surface, which act as electron donors or acceptors. Just like graphene, the variation of carrier concentration induced by adsorbates can be utilized to make highly sensitive sensors.[15] In this regards, 2D materials are very suitable for this purpose because of their large surface-to-volume ratio and high conductivities. Consequently, the effect of gas adsorption on the change in carriers concentrations can be maximized. Actually, it has demonstrated that single-layer $MoS_2$ based devices had a rapid and dramatic response to NO gas molecules but with an unstable current.[13] In order to interpret the result, it is highly desirable to elucidate the changes of electronic structure of $MoS_2$ after the adsorption of gas molecules, especially the charge transfer mechanism. Investigations of several adatom adsorption on $MoS_2$ monolayer have been shown that there is a considerable charge transfer between adatoms and $MoS_2$ layer.[16] However, the charge transfer between gas adsorbate and $MoS_2$ is still elusive.

To fully exploit the possibilities of gas sensors based on single layer MoS$_2$, it is important to understand the interaction between the MoS$_2$ surface and the adsorbate molecules. In this paper, we present a comprehensive study of the adsorption of various gas molecules on MoS$_2$ monolayer including CO, CO$_2$, NO, NO$_2$, and NH$_3$. The orientations and binding energies of these molecules on the surface are determined. The analysis of electronic structure of adsorbate and monolayer reveals that there is charge transfer occurs, which should have profound implications for its applications in gas sensors.

## II. Computational methods

First-principles calculations are carried out based on DFT[17] with the projector-augmented wave (PAW)[18, 19] method, as implemented in the Vienna *ab initio* simulation package (VASP).[20, 21] The generalized gradient approximation (GGA) functional of Perdew, Burke and Ernzerhof (PBE)[22] is used to treat the exchange and correlation potentials. Valence electrons for Mo are generated in $4p^65s^14d^5$ configuration and the configuration $3s^23p^4$ is used for the generation of the valence electrons of S. It is well known that both local density approximation (LDA) and GGA cannot capture the vdW interactions in weakly bonded systems such as gas adsorption on monolayer considered in this work. Therefore, it is necessary to incorporate additional functional into standard DFT calculations in order to correctly account for the effect of vdW interactions. In this work, it is included with two approaches: one is the DFT-D2 method of Grimme[23] which adds a semiempirical pairwise force field to conventional DFT calculations, and the other is the vdw-DF approach which adds a nonlocal correlation functional that approximately account for dispersion interactions.[24, 25] Two vdw-DFT methods with exchange-correlation energy given by optPBE[26] and revPBE[25] are considered.

All calculations are performed with a 5×5×1 supercell of 1H-MoS$_2$ containing 75 atoms. A large spacing of 15 Å between two-dimensional single layers of MoS$_2$ was used to avoid interlayer interactions. The total energy was converged to better than 10 meV for a plane wave cutoff of 500 eV and 5×5×1 Monkhorst-Pack[27] k-point sampling for the Brillouin zone. For geometry relaxation we used the method of conjugate gradient energy minimization. The convergence criterion for energy is chosen to be $10^{-4}$ eV between two consecutive steps, and the maximum Hellmann-Feynman force exerting on each atom is less than 0.03 eV/Å upon ionic relaxation.

The adsorption of five different gas molecules including CO, $CO_2$, NO, $NO_2$ and $NH_3$ on $MoS_2$ monolayer are investigated. The adsorption energy of gas molecules on $MoS_2$ is defined as $E_{ad}=E_{Gas+MoS2}-E_{MoS2}-E_{Gas}$, where $E_{Gas+MoS2}$ is the energy of the optimized structure of gas adsorbed on $MoS_2$, $E_{MoS2}$ is the energy of the pristine $MoS_2$ single layer and $E_{Gas}$ is the energy of isolated gas molecule. According to this definition, a negative $E_{ad}$ value indicates that the adsorption of gas molecules on the surface of $MoS_2$ is energetically favorable.

**III. Results and discussions**

In order to validate both the pseudopotentials and the method used in this simulation, we first calculated the lattice constant of pristine $MoS_2$ sheet using a unit cell. The calculated value is 3.18 Å, which is in excellent agreement with previous theoretical results of 3.20 Å.[16]

The adsorption energies of various gas molecules considered in this work on $MoS_2$ are summarized in Table I. Since we are mainly concentrated on the influence of gas adsorption on the electronic structure of $MoS_2$ monolayer, we do not consider different orientations of adsorbed gas molecules. However, the analysis of electronic structure is practically independent of the orientations and adsorption sites as pointed out previously. It can be seen that the adsorption energy of CO, $CO_2$ and $NH_3$ are rather small, suggesting a weak adsorption. On the other hand, the adsorption energy of NO and $NO_2$ are much large, which indicate a strong binding between these two molecules and $MoS_2$ layer. Here, it is useful to compare the adsorption energy of these molecules in graphene since graphene has been demonstrated to hold excellent chemical sensor properties. The $E_{ad}$ values are calculated to be 8-14 meV for CO and 15-31 meV for $NH_3$ on graphene based on GGA functional.[28] These results are much larger than those adsorbed on $MoS_2$ as indicated in Table I. However, the $E_{ad}$ values for NO and $NO_2$ on $MoS_2$ monolayer are totally different from those on graphene. For NO, the $E_{ad}$ values on $MoS_2$ is determined to be 35.8 meV which is larger than that on graphene with 11-29 meV, while $E_{ad}$ values for $NO_2$ on $MoS_2$ is much smaller than that on graphene (55-67 meV).[28, 29] Since a higher adsorption energy gives rise to a strong binding between adsorbate and the host, we can see that there is a stronger interaction between NO and $MoS_2$ monolayer compared to NO and graphene. In view of this conclusion, it is anticipated that the $MoS_2$ single layer should be more sensitive than graphene when used as NO detection device.

All these adsorption energies are significantly increased when the vdW interactions are included, as can be seen from Table I. However, there exist no detailed experimental data about the adsorption of gas molecules on $MoS_2$ monolayer so far. Based on previous calculations about the adsorption of cyclic molecules on $MoS_2$ basal plane,[30] the results obtained from optPBE-vdw should be more closer to experimental values. The adsorption energy of $NO_2$ on $MoS_2$ becomes the largest among the five molecules when vdw interactions are included, in contrast to conventional PBE results. The stronger binding between $NO_2$ and $MoS_2$ is in line with recent experimental and theoretical results.[31] In addition, the adsorption energy of $CO_2$ becomes lager and comparable to that of NO and $NO_2$ when the vdw interactions are included, indicating that vdw interaction dominates during the adsorption process.

Table I Adsorption energies (eV) of various gas molecules on $MoS_2$ monolayer determined from different methods.

| Molecules | PBE | DFT-D2 | optPBE-vdw | revPBE-vdw |
| --- | --- | --- | --- | --- |
| CO | -0.003 | -0.073 | -0.163 | -0.143 |
| $CO_2$ | -0.004 | -0.139 | -0.253 | -0.210 |
| $NH_3$ | -0.009 | -0.127 | -0.176 | -0.130 |
| NO | -0.066 | -0.153 | -0.254 | -0.239 |
| $NO_2$ | -0.036 | -0.138 | -0.287 | -0.241 |

The optimized adsorption configurations of various gas molecules on $MoS_2$ determined by PBE calculations are shown in Figure 1, in which the bond length of gas molecules and their nearest Mo or S atoms are indicated. It should be pointed out that the bond lengths of free gas molecules obtained are in agreement with experimental values. For example, the bond length of free CO is calculated as 1.14 Å, which is in line with 1.13 Å from experiment.[32]

For all the molecules, no chemical bonding is found and the entire bond is provided by weak vdW interactions. As a result, the bond length of gas molecules after adsorption is basically unaltered compared to the free molecules. The CO lies flat at about 3.50 Å above the basal plane and $CO_2$ locates 3.18 Å above $MoS_2$ sheet. The smaller distance of $CO_2$ indicates relative stronger interactions between them. For non-linear $NH_3$ and $NO_2$, the stable adsorption orientations are symmetric towards the sheet with the N atom located in the center. The stable configuration of NO is more closer to $MoS_2$ sheet with a tilt structure. Since the inclusion of vdw interactions changes the stable configurations very little, only the stable configurations obtained from PBE calculations are given.

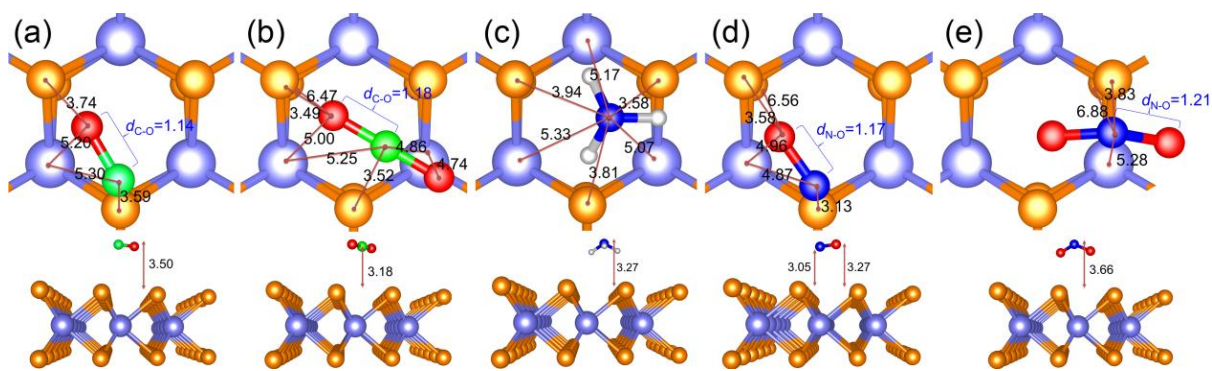

**Figure 1**. The most stable configurations of $MoS_2$ monolayer absorbed with (a) CO, (b) $CO_2$, (c) $NH_3$, (d) NO and (e) $NO_2$ from the top view (first row) and side view (second row) determined by PBE calculations. Mo atoms are in iceblue, and S atoms are in orange. C, O, N and H atoms in molecules are represented by green, red, blue and white balls, respectively. The bond lengths with units of angstrom are labeled around gas molecules.

The total density of states (DOS) of $MoS_2$ monolayer adsorbed with non-paramagnetic CO, $CO_2$ and $NH_3$ are presented in Figure 2. It can be seen from the DOS of pure $MoS_2$ that it has a finite bandgap of 1.52 eV, in accordance with previous results of 1.58 eV.[16] The DOS for either the valence or conduction bands of MoS$_2$ is not significantly influenced upon CO and NH$_3$ adsorption, which is consistent with their small adsorption energies indicated in Table I. The adsorption of CO$_2$ induces several distinct states at the lower-lying valence bands in the energy range of -8 eV to -6 eV and higher-lying conduction bands in the energy range of 4 eV to 6 eV. However, none of these three molecules adsorption can lead to a noticeable modification of DOS near the Fermi level. As a result, it can be concluded that the adsorption of CO and NH$_3$ does not have a substantial effect on the electronic structures of MoS$_2$.

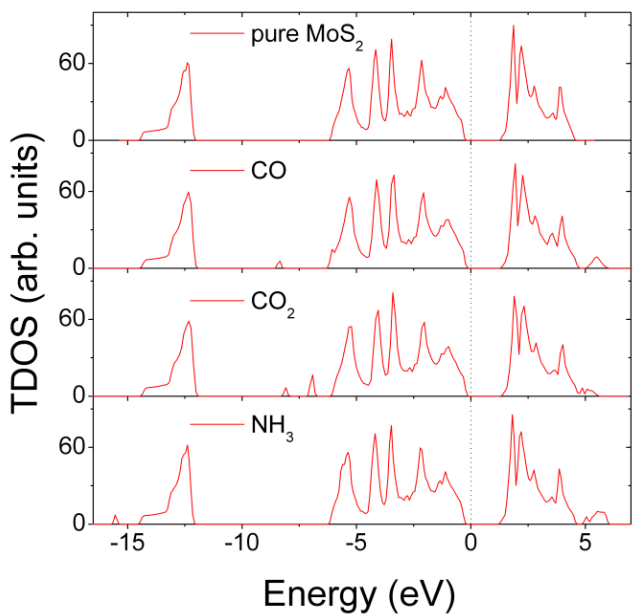

**Figure 2.** Total DOS of MoS$_2$ monolayer after adsorption of CO, CO$_2$ and NH$_3$. The Fermi level is indicated by the dotted line.

As found in graphene,[29] the adsorption with paramagnetic molecules such as NO and NO$_2$ can lead to a larger doping of MoS$_2$, which is manifested in their higher adsorption energy. The spin polarized DOS of NO and NO$_2$ adsorbed on MoS$_2$ are displayed in Figure 3, with the distribution of spin density included in the insets. It can be seen that the magnetization is mainly located on the NO or NO$_2$ molecules for both cases.

For NO adsorbate, the total magnetic moment of MoS$_2$ and adsorbate is 0.998 $\mu_B$. The Fermi level is shifted to conduction band which can be identified by DOS plot, indicating a n-type doping effect. In this case, charge transfer from NO to $MoS_2$ occurs. The differences around Fermi level are mainly attributed to the N(*p*) orbitals according to projected DOS analysis, which results in a DOS peak for spin-up electrons. Adsorption of $NO_2$ on $MoS_2$ leads to a magnetic moment of 0.998 $\mu_B$. It is seen that the Fermi level is downshifted to valence band, indicating a hole doping for $MoS_2$. The DOS peak of spin-down electrons is caused mainly by N(*p*) orbitals.

It is inferred experimentally that there is charge transfer resulted from electron-withdrawing $NO_2$ molecules adsorbed on $MoS_2$ film that leads to the increase of conductance in p-type transistors. Our charge analysis based on Bader charges[33, 34] reveals that the adsorbed $NO_2$ accepts about 0.034e$^-$ from the $MoS_2$ monolayer, indicating that $NO_2$ works as an acceptor. The charge transfer is expected to induce changes on the conductivity of the system.

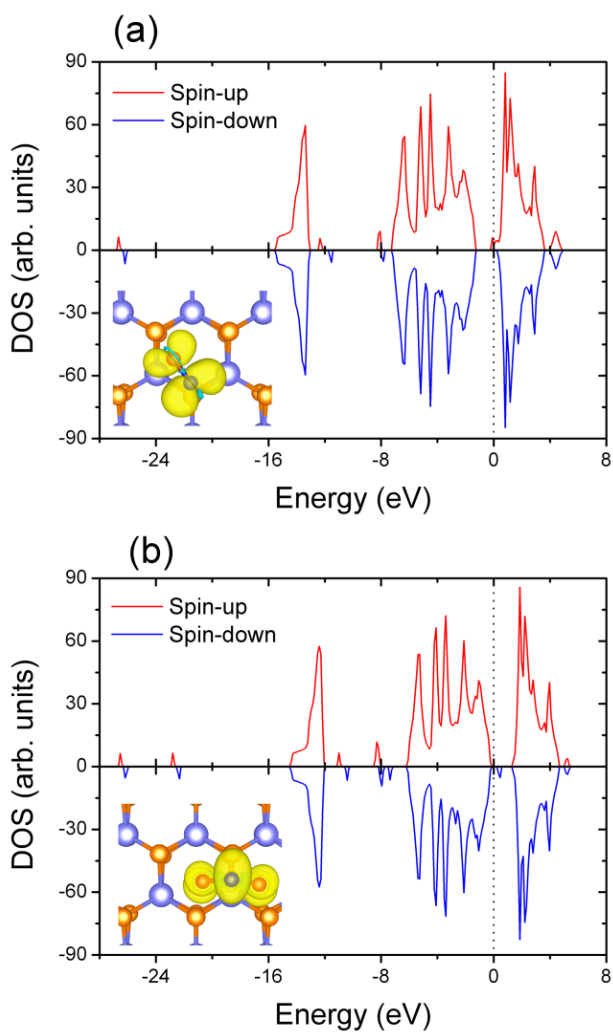

**Figure 3.** Spin-polarized density of states of (a) NO and (b) $NO_2$ adsorbed on $MoS_2$. The spin density plot are shown in the insets with the isovalue is set to be ±0.002 a.u.

To further elucidate the charge transfer between gas adsorbates and $MoS_2$ monolayer, we provide in Fig. 4 the isosurface plot of electron charge density difference for these gas molecules. Here the charge difference is obtained by subtracting the electron densities of non-interacting components (gas and $MoS_2$ monolayer) from the charge density of the gas-$MoS_2$ system while retaining the same atomic positions of the components and the gas-$MoS_2$ system. It is indicated that there is a charge accumulation on $MoS_2$ for CO, $CO_2$, $NH_3$ and NO adsorbate, suggesting that their charge-donor characteristic. On the other hand, $NO_2$ acts as electron acceptor and accepts electrons from $MoS_2$ as evidenced by the charge depletion region near the $MoS_2$ sheet.

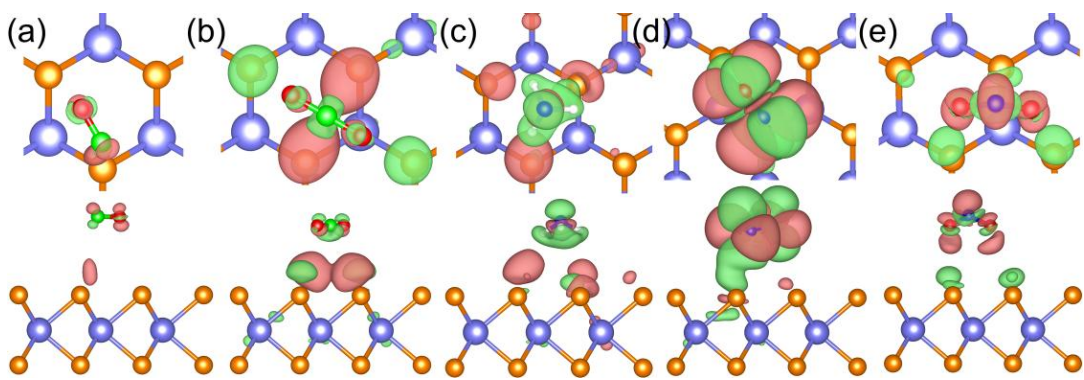

**Figure 4.** Isosurface plot of the electron charge density difference for (a) CO, (b) $CO_2$, (c) $NH_3$, (d) NO and (e) $NO_2$ on $MoS_2$ monolayer with the isovalue of ±0.0002 a.u (top view and side view are provided in first row and second row, respectively). The charge accumulation is represented in pink and charge depletion is in lime, respectively.

It is demonstrated that there is large charge transfer occurs between NO and $MoS_2$, which is also reflected by its high adsorption energy as describe above. Consistent with DOS plot as shown in Fig. 3, NO denotes electrons and acts as electron donor while $NO_2$ acts as an electron acceptor and accepts electron from the $MoS_2$ monolayer. The charge transfer between gas molecules and $MoS_2$ sheet can lead to the variation of its resistance when it is exposed to these gases. For example, charge depletion in MoS$_2$ sheet reduces the number of electrons in the monolayer and thus lowers its charge carriers and enhances its resistance.

In order to gain more insights into the mechanism of adsorption and site preference, we have examined the evolution of the local electronic structure of the NO and the surface upon adsorption. The obtained projected density of states (PDOS) are demonstrated in Fig. 5, in which the PDOS of NO and its nearest S and Mo atoms in the stable adsorption site are analyzed. Note that the molecular orbitals (MOs) of free NO termed as 3σ, 4σ, 5σ, 1π and 2π* are labeled in Fig. 5(a), respectively. It can be seen that the DOS peak of NO are all narrowed upon adsorption due to the transfer of electrons from NO to MoS$_2$. Closer inspection reveals that the 1π, 5σ and 2π* orbitals are the most affected orbitals by adsorption with a large downshift. In particular, the 2π* orbitals are found to be split into two separated peaks. There are no new orbital occupations in the DOS of NO after adsorption, indicating no charge transfer from electronic states of MoS$_2$ to NO molecules.

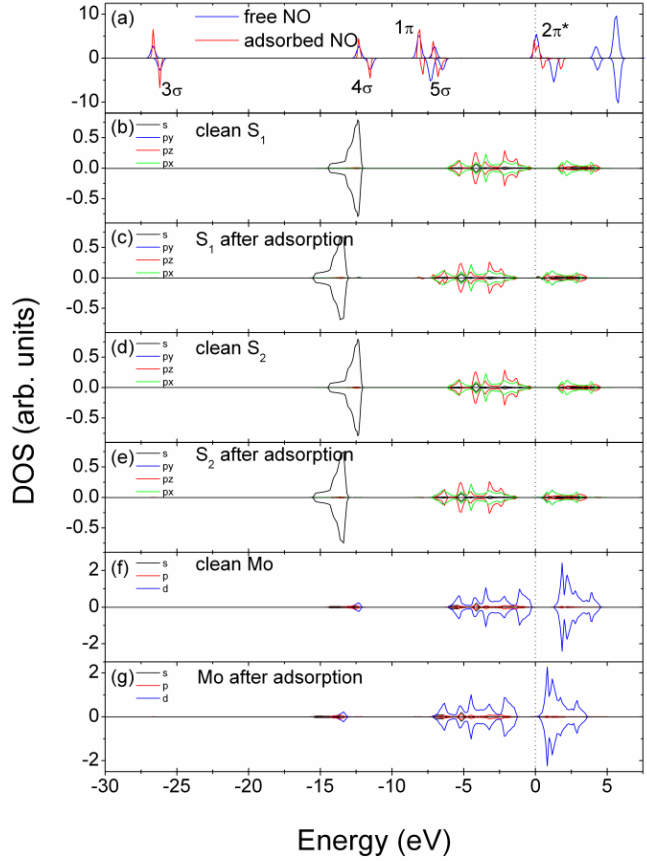

**Figure 5.** (a) PDOS for the NO molecule before and after the adsorption. (b)-(g) PDOS for the two S and one Mo atoms nearest to adsorbed NO before and after the adsorption. The Fermi energy is set to zero denoted by dotted line.

The DOS of nearest S and Mo are shifted to lower energies during adsorption process, which implies that they accept electrons from bonding molecular orbitals of NO. However, the DOS profiles are generally not affected by the adsorption of NO, suggesting the weak interactions between them. In the vicinity of Fermi level, there is a pronounced hybridization between $2\pi^*$ orbitals of NO and $d$ orbitals of Mo atoms in the surface, which manifests in the large adsorption energy as indicated in Table I.

The change in the DOS of $MoS_2$ in the band gap region upon NO adsorption implies a sensitive response of $MoS_2$ to NO. On the other hand, the weak interaction between NO and $MoS_2$ monolayer means that NO can fast desorption from the sheet, which accounts for experimental observation that FET based on $MoS_2$ shows a rapid and dramatic response upon exposure to NO but with an unstable current.[13]

## IV. Summary

First-principles calculations are carried out to investigate the adsorption of various gas molecules on $MoS_2$ in order to fully exploit the applications of $MoS_2$ gas sensors. With vdw interactions included, the binding of NO and $NO_2$ with $MoS_2$ is shown to be the strongest among the gas molecules considered, suggesting that $MoS_2$ is more sensitive to these two gases. The change in the electronic structure and charge transfer induced by gas adsorption is found to be responsible for the strong binding observed in NO and $NO_2$. Our results thus provide a theoretical basis for the application of MoS2 monolayer as gas sensors for important polluting gases such as NO and $NO_2$ in air.